\documentclass[preprint,3p,11pt]{elsarticleGAC}

\usepackage{etoolbox}
\makeatletter
\patchcmd{\ps@pprintTitle}{\footnotesize\itshape
       Preprint submitted to \ifx\@journal\@empty Elsevier
       \else\@journal\fi\hfill\today}{\relax}{}{}
\makeatother

\usepackage{amsmath}
\usepackage{amsfonts}
\usepackage{amssymb}
\usepackage{times}
\usepackage{graphicx}

\begin{document}

\title{Born's prophecy leaves no space for quantum gravity}

\author{Giovanni AMELINO-CAMELIA\\
{\small{\it{Dipartimento di Fisica, Sapienza Universit\`a di Roma and INFN, Sez.~Roma1, P.le A. Moro 2, 00185 Roma, EU}}}}
%

\vskip -0.9cm

\begin{abstract}
I stress that spacetime is a redundant abstraction, since
describing the physical content of
all so-called ``space-time measurements" only requires timing (by a physical/material clock)
of particle detections (at a physical/material detector).
It is interesting then to establish which aspects of our current theories
afford us the convenient abstraction of a spacetime.
I emphasize the role played by the assumed triviality
of the geometry of momentum space, which makes room for an observer-independent notion
of locality.
This is relevant for
some recent studies of the quantum-gravity problem that
stumbled upon hints
of a nontrivial geometry of momentum space, something which had been strikingly envisaged
for quantum gravity already in 1938 by Max Born.
If indeed momentum space has nontrivial geometry then
the abstraction of a spacetime becomes more evidently redundant and
less convenient: one may still abstract a spacetime but only allowing
for the possibility of a relativity of spacetime locality.
I also provide some examples of how all this could affect our attitude toward the quantum-gravity
problem,
including some for the program of emergent gravity
and emergent spacetime and an indication of triviality of the holographic
description of black holes.
And in order to give an illustrative example of possible logical path for the ``disappearance
of spacetime" I rely on formulas inspired by the $\kappa$-Poincar\'e framework.
\end{abstract}

\maketitle

\baselineskip11pt plus .5pt minus .5pt

\section{Redundancy of spacetime and primitive measurements as timed particle detections}	
It is easy to see~\cite{principle,grf2nd} that the most
basic (``primitive") ingredients of all of our so-called ``spacetime measurements"
are particle detections, possibly (but not necessarily) also recording the
time of the detections, according to a local physical clock, and
also recording
the momentum of the particle, if the detector is, {\it e.g.}, a calorimeter.

This is particularly evident when we perform ``spacetime measurements" in astrophysics,
such as those using observations of particles from a source to "localize" that source.
But it is actually no less true for all other "spacetime measurements", including for example
those involving the use of a ruler or a tape meter.

Evidently spacetime's popularity as a notion used in physics
 must be the result of its being a very convenient abstraction, and most of these notes will
be about recognizing in which ways this abstraction is convenient and
recognizing which aspects of our current ``understanding"/formulation of the laws
of physics affords us this abstraction.

\subsection{Emission-detection type and wasteful redundancies}
Having the objective of keeping these notes reasonably short,
I shall illustrate some concepts through single simple
examples of their applicability.
In this subsection I work in a 1+1-dimensional spacetime
and consider the case which in our presently conventional
description corresponds to flat background Minkowski spacetime.
The measurement of emission-detection type which I consider
involves an ``observer" Alice, which is an emitter, and an ``observer" Bob which is a detector,
and specifically the case where Alice and Bob are in relative rest at a distance $L$
from each other, with two synchronized clocks located one at Alice and one at Bob.
Alice might write the following equation
to describe the ``trajectory in spacetime" (the worldline) of a certain specific
particle:
\begin{equation}
x=ct
\label{worldline}
\end{equation}
Of the infinitely many "truths" codified in this equation
only two facts are established experimentally: the particle is emitted at Alice
at time $t=0$ of Alice's clock and the particle is detected at Bob at time $t=x/c$
of Bob's clock.

We do need some notion of distance between the detectors
actually at work and a notion of synchronization
(or specifiable lack of synchronization) of clocks at those detectors,
but the spacetime picture is redundant.

This is particularly clear in relativistic theories, and plays a role in the
relationship between active and passive relativistic transformations.
There the redundancy is in characterizations such as ``the particle was at a distance of 10 meters
from Alice, where Bob is", which evidently can be fully replaced by ``the particle was at Bob".
In the case discussed above a particle emitted at Alice at $t=0$
is detected at Bob at $t=L/c$, with the relationship between Alice and Bob such that
it is given by a translation transformation with translation parameter $L$.\\
This is for pure translations, but the generalization that includes boosts is
equally trivial.

And notice that the redundancy of spacetime becomes more manifest in cases where
  quantum-mechanical effects
 are relevant. If there is a ``double slit" between Alice and Bob
 the redundant spacetime description must be creatively adapted, but
 actually nothing changes about
 the scientific part of the description (with the rather marginal modification that some detection
 probabilities which could be idealized as $100\%$ within classical mechanics are often
 unavoidably smaller than $100\%$ within quantum mechanics).

\subsection{Bubble-chamber type and convenient redundancies}
If all we had were emission-detection-type measurements the abstraction of a spacetime
would have probably never been made, since in those measurements it is
{\underline{wastefully}} redundant.
But many of our ``spacetime measurements", particularly in the classical regime,
are not of emission-detection type but rather
of a type that could be described as a sequence of measurements of localization
of the same particle (or body), in which case the abstraction of a spacetime
is {\underline{usefully}} redundant.
Again within the vast collection of such measurements (which of course
include, {\it e.g.}, those performed when we watch a game of soccer)
I choose to focus on only one, which I label ``bubble-chamber type".

Of course also for a bubble-chamber setup the primitive measurements
are timed particle detections, which in most
bubble-chamber setups will be photons detected by our eyes (or by a photographic camera).
That sequence of photons however proves to be reliably describable in terms
of the abstraction of positions for certain bubbles, and in turn the collection
of bubbles allows us to infer reliably the abstraction of a ``trajectory" for a charged
microscopic particle. Evidently here too the spacetime abstraction is redundant:
we could describe a bubble-chamber measurement exclusively in terms
of timed detections of many photons. But the advantages of adopting the
spacetime-trajectory abstraction in such cases are undeniable.

Among these ``bubble-chamber-type measurements" I should also mention
the ones relevant for many studies in astrophysics: when a star bursts or explodes
we detect bunches of photons (occasionally also neutrinos, and maybe one day gravitons).
The discussion of the physics content of these
measurements could in principle be limited to these timed particle detections,
but it is evidently advantageous to recognize that in these instances the
sequence of particle detections can be organized into the abstraction
of a localized explosion ``far away in spacetime".

\subsection{Spacetime and the ether}

\noindent
What is then spacetime?\\
I shall leave most aspects of this question
to the appetites of philosophers, who have license to debate (endlessly of course)
about things that ``exist but are not observed".
For the science of physics one can confine the discussion of this question
to two simple obervations:\\
(i) the fact
that we can add reference to a ``spacetime" without adding any new item
to our list of measurement procedures (still only timed particle detections)
is a complete proof of the redundancy of spacetime in science;\\
(ii) but, while awareness of its redundancy may at some point be valuable,
 the abstracted notion of spacetime
is in some sense tangibly useful, at least in our current theories
({\it i.e.} at least within the confines of the regimes of physics
so far explored),
as stressed in the previous section.
 And as long as
this is the situation there would be no reason of course
for us to change
the way to do physics, using the spacetime notion.

\newpage

It will be clear from these notes that
it seems to me the status of spacetime in our current theories shares
some aspects of the status
of the ether at the beginning of the 20th century.
I find evidence on the internet of Poincar\'e stating, in 1889,
that
{\quotation{{\it ``Whether the ether exists or not matters little - let us leave that to the metaphysicians; what is essential for us is, that everything happens as if it existed, and that this hypothesis is found to be suitable for the explanation of phenomena. After all, have we any other reason for believing in the existence of material objects? That, too, is only a convenient hypothesis; only, it will never cease to be so, while some day, no doubt, the ether will be thrown aside as useless."}}}

\vskip 0.4cm 

Still focusing, for no good reason, on Poincar\'e, and offering something else of indirect
relevance to my thesis,
I must highlight
his rather unfortunate encounters (present in several of his works both before 1905 and even after~\cite{poinca1908} 1905)
 with the notion of ``true time of resting clocks in the ether",
 used in descriptions such as this:
{\quotation{{\it``Pour que la compensation se fasse, il faut rapporter les phenomenes, non
pas au temps vrai $t$, mais a un certain temps local $t'$ defini de la facon suivante.
Je suppose que des observateurs places en differents points, reglent leurs montres
a l'aide de signaux lumineux; qu'ils cherchent a corriger ces signaux du temps
de la transmission, mais qu'ignorant le mouvement de translation dont ils sont
animes et croyant par consequent que les signaux se transmettent egalement
vite dans les deux sens, ils se bornent a croiser les observations en envoyant un
signal de A en B, puis un autre de B en A. Le temps local $t'$ est le temps
marque par les montres ainsi reglees. Si alors $c$ est la vitesse de la lumiere, et $v$
la translation de la terre que je suppose parallele a l'axe des $x$ positifs, on aura:}
$t' =t -vx/c^2$."}}

\subsection{More on primitive measurements}
I have so far used the expression ``primitive measurement" assuming my readers
will share an intuitive perception of what I mean by it. I am unable to go much beyond
that, but in closing this section I can attempt to give more structure to the intuition.

The most primitive measurements are the ones by our ``resident devices",
the monolithic measurement procedures associated with
smell, touch, taste, sight  and sound. Confirming my thesis, these are most evidently
timed particle detections at some macroscopic detectors.
Timing at this stage is provided by our resident clock, which is evidently there
(though I do not understand much about it) and has to do with ``memory"
(a very low quality but rather robust clock).
No other measurement could possibly be more primitive than those of our resident
devices, but since those devices have very poor resolution, it is pointless to contemplate
the level of fundamental physics that codifies exclusively observations by our resident devices.

Indeed because of this issue about the resolution of our resident devices, I am
including among primitive measurements also those which involve a single step of inference.
For this I consider those who can at least in principle be described as measurements
using devices which can be operated as {\underline{sealed boxes}} with a digital display
on the surface of the box. Among these there is evidently no spacetime-measurement device.
And the two examples I do see clearly are clocks and calorimeters.
One can easily imagine adopting as standard for timing a certain specific class of identical
sealed boxes with digital displays. The only stage of inference
involved in making use of such clocks comes in when we read the clock,
a single stage of inference involving particle
detections.
Similarly primitive are particle detections with {\underline{energy}} (momentum)
determination by a ``generalized calorimeter" (an actual calorimeter or something that can serve
the same purposes of a calorimeter).
Again it is not hard to imagine adopting as standard for such devices
 a certain specific class of identical
sealed boxes with digital displays. The only stage of inference
involved in making use of such ``calorimeters" comes in when we read the display, again
a single stage of inference involving particle
detections.

It should be easily appreciated that no spacetime measurement (such as measurements of distances
or lengths) can aspire\footnote{Part (but not all!) of this originates from the fact
that spacetime measurements are typically nonlocal. The easiest examples of this are of the type of
measurements of lengths, where the nonlocality of the measurement procedures
is very explicit. Other examples are more subtle, as in the case of velocity measurements: velocity
is formally a ``local observable" but its measurement involves necessarily
a nonlocal procedure.} to be as primitive as the particle detections by
our  ``resident devices".
And no sealed-box setup is available for spacetime measurements: I can leave a calorimeter
unattended for all of its useful time of activity, with only a human in charge of reading
 the digital display,
and still have a sequence of measurements, whereas there is no analogous sealed-box procedure
that will provide us length measurements.

Of course, some of the procedures we conventionally label energy (/momentum) measurements
are not primitive: we can use the available theoretical framework to infer an
energy (/momentum) measurement through analyses that rely on several layers of inference.
But this is evidently
irrelevant for my thesis. There is a quality that timed particle detections with
momentum determination {\underline{can}} have which spacetime measurements {\underline{cannot}} have.

Actually it should be clear that most measurements in physics are far removed from
the status of being primitive: in particular both particle detections and spacetime measurements
usually are performed relying on several layers of inference. As long as we trust our
inferences (as long as the assumption that the inferences are trustworthy holds up)
worrying about these layers of inference falls in the ``sex-of-the-angels category".
But in cases, such as quantum-gravity research, where we look for a new theory paradigm
and we may fear some of these inferences to become untrustworthy, awareness of
the different layers of inference can be important.

In particular I am here arguing
it is important to notice that our most primitive measurements all
are timed particle detections.
And I feel that at some stage it may become also important to distinguish finely
among the amount of inference (the number of layers of inference) needed by some specific
measurement procedures: comparing two non-primitive measurements we would typically find
that one is more primitive (less non-primitive) than the other.
As an example of this I can compare two spacetime measurements performed on a planar capacitor:
a measurement of the area of the plates and a measurement of the distance between the plates.
One can measure the area of the plates from the capacity of the capacitor {\underline{if}}
the distance between the plates is determined experimentally independently.
So the results of this specific type of area measurement is less primitive
than the result of the distance measurement (measurement of the distance between the plates)
that it uses.

And it should be also evident that the facts on which our abstraction of a macroscopic spacetime
rests involve several layers of inference. The way by which in astrophysics we characterize
operatively the position in spacetime of a distant burst of particles is indeed a very
clear illustration of these layers of inference. And this is not of mere academic interest,
since it may well be that the first evidence of the need of adopting a new theory paradigm
for quantum gravity might come~\cite{myLRR} from anomalous properties of this sort of observations
of distant bursting sources.

\section{The role of absolute locality}


\subsection{Absolute locality and translation transformations}
From the perspective I am adopting
it is interesting then to ask which structural properties
of our presently used theories render the abstraction of spacetime
so convenient, and if we can see on the horizon any reason for these structural
properties to be lost.

As recently emphasized in Refs.~\cite{principle,grf2nd}, a key aspect of our current theories
which plays a role in the convenience of the abstraction of a spacetime is ``absolute locality".
For example when an observer $O$ detects particles from a distant explosion, and
uses those particle detections as identification of a ``distant point in spacetime",
then, and this is absolute locality, all other observers agree that a point-like source
exploded. The point of spacetime obtained by abstraction from this procedure of course
carries different coordinates in the different reference frames of the two observers,
but the laws of transformation among observers renders this all intelligible, ensuring
that the two observers ``leave in the same spacetime", that they simply give different
coordinates to the same objective (abstraction of) spacetime point.

It was recognized in Refs.~\cite{principle,grf2nd} that a key role in establishing
this absolute locality is played by the triviality of our current description of
translation transformations. Take for example
 a distant 3-particle interaction $e_{k \oplus p \oplus q}$,
 which could be the decay of a pion of momentum $k_\mu$ into a muon of momentum $p_\mu$
and a neutrino of momentum $q_\mu$.
From a locality perspective this must be viewed as a coincidence of 3 events:
the disappearance of the pion and the appearances of the muon and the neutrino.
According to an observer $O$ the interaction
occurs at the spacetime point with coordinates $x^\mu$ ,
which means that at the interaction the coordinates
of the three particles coincide $x^\mu_{(1)} = x^\mu_{(2)} = x^\mu_{(3)}=x^\mu$.
In our current theories, because they indeed have absolute locality,
another observer $O'$, at distance $b^\mu$ from $O$, will
 also describe the same 3-particle interaction as a coincidence of 3 events,
with $x^{\prime\,\mu}_{(1)} = x^{\prime\,\mu}_{(2)} = x^{\prime\,\mu}_{(3)}= x^{\prime\,\mu}$.
This is because of the triviality of translation transformations:
$$x^{\prime\,\mu}_{(n)} = x^{\mu} + \{ x^\mu_{(n)} , b^\nu { P }^{tot}_\nu \}
= x^{\mu}_{(n)} + \{ x^\mu_{(n)} , b^\nu p_{(n)\,\nu} \}
= x^{\mu}_{(n)} + b^\mu$$
Since our current description of translation transformations works this way it is inevitable
that a coincidence of events 
for observer $O$ will also be described as a coincidence of events by another
observer, $O'$, at rest with respect to $O$ at distance $b^\mu$ from $O$.
But in Planck-scale theories with deformed translation transformations we must then
be prepared to loose absolute locality (more later and in Refs.~\cite{principle,grf2nd}).

\subsection{Absolute locality and boost transformations}
Before I discuss a possible path (``Born's prophecy") for how absolute
locality might be lost in the quantum-gravity realm,
let me first generalize the characterization of absolute locality given in
Refs.~\cite{principle,grf2nd}, from the case of translation transformations there considered
to the more general case of transformations involving translations and/or boosts.
Also the simplicity
of our present description of boost transformations is responsible for
the fact that we are afforded the luxury of absolute locality.

To see this let us consider again
the example of 3-particle interaction $e_{k \oplus p \oplus q}$,
 decay of a pion into a muon
and a neutrino.
As done above, for observer $O$ the interaction
occurs at the spacetime point with coordinates $x^\mu_{(1)} = x^\mu_{(2)} = x^\mu_{(3)}=x^\mu$.
Rather than taking $O'$ as a distant observer in relative rest with respect to $O$,
I now take $O'$ as an observer purely boosted with respect to $O$.
And of course once again, since our current theories do afford us absolute locality,
also $O'$, boosted by rapidity $\xi^j$ with respect to $O$, will
 describe the  3-particle interaction as a coincidence of  events,
with $x^{\prime\,\mu}_{(1)} = x^{\prime\,\mu}_{(2)} = x^{\prime\,\mu}_{(3)}= x^{\prime\,\mu}$.
This is because of the triviality of our boost transformations,
which (focusing for brevity on the 1+1-dimensional case) take the form
$$x^{\prime\,1}_{(n)} = x^{1} + \{ x^1_{(n)} , \xi {N}^{tot} \}
= x^{1}_{(n)} + \{ x^1_{(n)} , \xi N_{(n)} \}
= x^{1}_{(n)} - \xi x^0_{(n)}$$
$$x^{\prime\,0}_{(n)} = x^{0} + \{ x^0_{(n)} , \xi {N}^{tot} \}
= x^{0}_{(n)} + \{ x^0_{(n)} , \xi N_{(n)} \}
= x^{0}_{(n)} - \xi x^1_{(n)}$$
As I shall stress in the following
some aspects of the study of the quantum-gravity problem may suggest
a path for loosing absolute locality in ways that also involve
a modification of this state of affairs for boosts.

\section{Born's prophecy and a quantum-gravity path to relative-locality momentum spaces}
Absolute locality plays a key role in rendering the abstraction of a spacetime
convenient. It ensures that ``we all live in the same spacetime"~\cite{grf2nd},
which is the main aspect of convenience of the spacetime abstraction:
if we remove absolute locality then for example a point-like distant source
for observer Alice (say, a gamma-ray burst) would not be a point-like source
for some observer Bob; points of Alice's spacetime would not be points of Bob's spacetime
and vice versa; Alice and Bob would not be different observers in the same spacetime,
but rather different observers, each abstracting a different spacetime.\\
This is scary and exciting. Could we actually loose absolute locality?\\
My next task is to show that some aspects of the quantum-gravity problem suggest
that we might loose it!\\
And remarkably it may be connected with a speculation made by Max Born already
in 1938~\cite{born1938}, which suggests that
 in order to have a consistent quantum-gravity theory one should be able to accommodate
 a curvature for momentum space.

I call this ``Born's prophecy" because it went largely unnoticed for several decades
(with the exception of a few isolated cases such as Ref.~\cite{golfand}).
But in recent times several different realizations of the idea of a curved momentum space
have surfaced in the quantum-gravity
literature~\cite{majidlectnotes,girelliCURVATURE,changMINIC,schullerCURVATURE}.
In this section I draw what I see as the most compelling path from the structure of
the quantum-gravity problem to the possibility of fulfilling Born's prophecy
and loosing absolute locality.

\subsection{From minimum wavelength to DSR}
The structure of the quantum-gravity problem is such that we are led to expect
that the Planck length (or a length scale not too different from the Planck length)
should play some role in the short-distance structure of spacetime.
I shall not here review the very extensive literature on this, but my readers can find
in Ref.~\cite{garay} a good effort of reviewing the status of the relevant debate
until the mid 1990s (a more recent perspective is in Ref.~\cite{sabineMINLENGTH}).
The roles attributed to the Planck length include:
minimum allowed value for wavelengths,
minimum allowed value for length,
and minimum allowed value for the uncertainty in length measurement.

Remarkably, all this was being contemplated
without appreciating that there could and perhaps
should be implications for Lorentz symmetry.
But if any of these speculations are correct
it is absolutely necessary to ask~\cite{dsr1,dsr2}
what would be the fate of Lorentz symmetry at the Planck scale.
Think for example of the statement ``the Planck length is the minimum allowed
value of wavelength": this is no less hostile~\cite{dsr1} to Lorentz symmetry
than a maximum-velocity principle is hostile to Galilean-boost symmetry.

An intense scrutiny of the fate of Lorentz symmetry at the Planck scale only
started with the first works~\cite{grbgac,gampul,mexweave}
 attracting sizable attention on the possibility
of Planck-scale modifications of the dispersion relation.
We now have several scenarios for accommodating notions such a Planck-length minimum wavelength
and/or Planck-scale-deformed dispersion relation
within a deformed-Lorentz-symmetry setup of DSR (``doubly-special relativity") type~\cite{dsr1,dsr2}.
Preparing for my objective of exposing the usefulness of $\kappa$-Poincar\'e-inspired
momentum space in the development of this literature, let me consider
the following on-shellness/dispersion relation (again, for simplicity, in 1+1-D spacetime)
\begin{equation}
\cosh (\ell m) = \cosh (\ell p_0) - \frac{\ell^2}{2} e^{-\ell p_0} p_1^2
\label{disprelkappa}
\end{equation}
which is invariant under the deformed boosts given by
\begin{equation}
[N, p_0] =  p_1 ~,~~~
[N, p_1] = \frac{e^{2 \ell p_0} - 1}{2\ell}  + \frac{\ell}{2} p_1^2~.
\label{boostkappa}
\end{equation}
Notice that according to (\ref{disprelkappa}) the  length scale $|\ell|$
is, for negative $\ell$,
the minimum wavelength (the inverse of the Planck length is the maximum momentum
of massless particles): this can be established by studying the action of boosts
on momentum space and finding~\cite{dsr1,dsr2} that, for negative $\ell$,
 this action on spatial momentum saturates
at $1/|\ell|$, and it can also be established~\cite{jurekdsr1} 
directly from the dispersion
relation (when $\ell$ is negative $p_1 \rightarrow 1/|\ell|$ as $p_0 \rightarrow \infty$).
Since the boosts (\ref{boostkappa}) ensure that the dispersion relation  (\ref{disprelkappa})
is observer independent, we have here a way to enforce as relativistic laws
both a modified
 dispersion relation and a principle of minimum wavelength.
In these DSR-relativistic theories the scale $\ell$ plays a role completely analogous
to the speed-of-light scale $c$ (not visible in my formulas because of the choice of units).

\subsection{From DSR to deformed momentum composition and boost-induced relative locality}
It was clear from the very beginning of studies of Planck-scale-deformed relativistic
kinematics~\cite{dsr1,dsr2} that deforming boost actions on momenta requires,
in order to preserve relativistic invariance, that also the law of composition
of momenta be deformed. The logical steps that lead to this conclusion are very general
and elementary, since they proceed just in the same way by which, taking as starting point
Galilean relativity, the introduction of an invariant speed scale (the speed-of-light
scale) requires a deformation of the law of composition
of velocities.
Let me illustrate this general feature within the context of
the setup for which I already specified the laws  (\ref{disprelkappa})
and (\ref{boostkappa}). It is evident that conservation laws based on the special-relativistic
composition law, $(p \oplus p^\prime)_\mu = p_\mu + p^\prime_\mu$,
such as the conservation law formally given by $p_\mu + p^\prime_\mu=0$,
are not covariant with respect to the boost action (\ref{boostkappa}); in particular, one
 finds
\begin{eqnarray}
&& [N_{[p]}+ N_{[p^\prime]},
p_1 + p^\prime_1]
 = \left(\frac{e^{2 \ell p_0 } -1}{2\ell}  + \frac{\ell}{2} p_1^2 \right)
+ \left(\frac{e^{2 \ell p^{\prime}_0 } -1}{2\ell}  + \frac{\ell}{2} p^{\prime \, 2}_1 \right)
\label{covathreeunoHHppkALL}
\end{eqnarray}
which does not vanish enforcing on the right-hand side the conservation
law\footnote{In these notes, for brevity, I discuss processes handling all particles as incoming.
 For example, momentum conservation in propagation, when the composition of momentum is trivial,
 is written as $p+p^\prime=0$ with, say, $p=p_{incoming}$ and $p^\prime=-p_{outgoing}$.
 An example with deformed law of composition of momenta $p \oplus k$, would be
 momentum conservation in propagation
 written as $p \oplus p^\prime=0$ with, say, $p=p_{incoming}$ and $p^\prime= \ominus p_{outgoing}$
 (where $\ominus$ denotes the ``antipode" such that $p \oplus [\ominus p] =0$).} itself, $p_\mu + p^\prime_\mu=0$.

But relativistic covariance of the conservation laws can be reinstated
by introducing suitably nontrivial~\cite{dsr1}
composition of momenta (and possibly suitably nontrivial
composition of boosts, see later) such that
$$[N_{[p]} \oplus^N_\ell N_{[p^\prime]},
(p \oplus_\ell p^\prime)_\mu]\Big|_{(p \oplus_\ell p^\prime)_\mu=0} =0$$
The list of setups that satisfy this criterion is at this point very large.
I'll discuss one explicitly later in this section, and other examples are provided in
Refs.~\cite{dsr1,judesvisser,frandar,leeDSRprd,goldenrule,hopfhopf} and references therein.

DSR-deformed Lorentz transformations also have another consequence, no less dramatic
then the deformation of the law of composition of momenta, and particularly relevant
for the subject of these notes: relative locality (see Fig.~1).

\begin{figure}[h!]
\includegraphics[width=0.99 \textwidth]{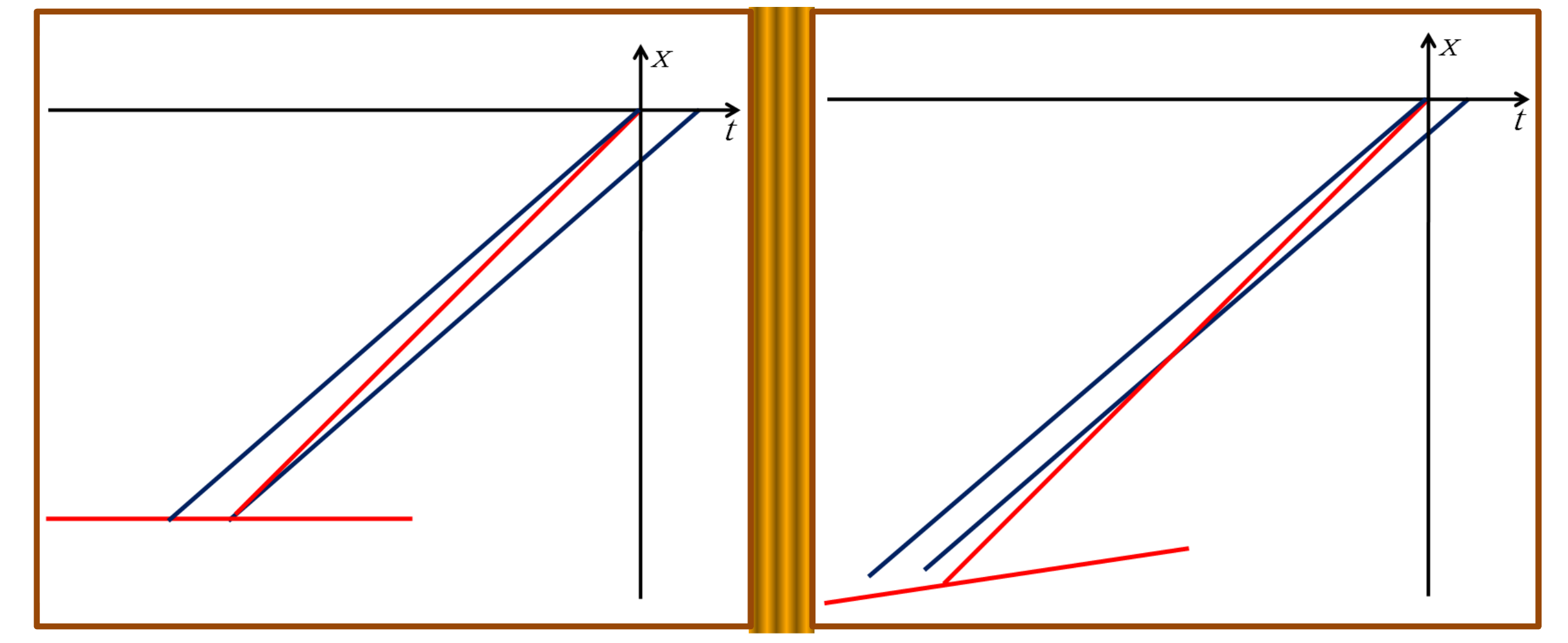}
\caption{I here illustrate a feature of relative spacetime locality
that follows from certain fashionable DSR-deformations of Lorentz boosts, as shown
in Ref.~[22].
The left panel shows an idealized case in which Alice's description has a coincidence of events,
 which she observes (in the origin
of here reference frame), and two distant coincidences of events, which she infers.
The right panel is for the description by Bob, purely boosted with respect to Alice,
of the same idealized situation: also Bob observes the
coincidence of nearby events, but for Bob the distant events are not exactly coincident.
Red lines are for low-energy worldlines, unaffected by the deformation, while blue lines
are for particles of energy high enough to be tangibly affected
(within the resolution available experimentally) by the deformation.}
\end{figure}

As first noticed in Ref.~\cite{bob},
typically DSR-deformed boosts affect\footnote{In retrospect, the relativity of locality of distant
coincidences of events first described in Ref.~\cite{bob} can be viewed as the answer to some
puzzles for absolute locality of distant events which had been noticed
in Refs.~\cite{gacIJMPdsrREV,Schutzhold:2003yp,Arzano:2003da,grilloSTDSR,dedeo,Hossenfelder:2010tm}.
A useful complementary way to characterize the presence of relativity of spacetime locality
is obtained, following Ref.~\cite{leeLIMITATIONS}, by establishing the finiteness
of the range of
distances and boosts from a given observer within which (for given value of the relevant
experimental resolution) one could still effectively rely on absolute locality.}
the description of coincidences of distant events.
For simplicity take the case of an observer Alice who is local witness of
a pion decay (a pion decay occurs in the origin of her coordinate system)
and also infers an ordinary coincidence of events characterizing a distant pion decay
(detects the decay products of a pion decay which occurred
far from her origin, but from the spatial-momentum components and timing of those detections
Alice infers a distant coincidence between the disappearance of the pion and the appearance
of, say, the muon and the neutrino).
It follows from the structure of typical DSR-deformed boosts that~\cite{bob}
an observer Bob purely boosted with respect to Alice will still necessarily describe
as ``local", {\it i.e.} as a coincidence of events,
 the pion decay in the origin of Alice (which is also the origin of Bob, since
they are related by a pure boost),
but instead in such a situation if Alice inferred a coincidence of events for the
distant pion decay then Bob will necessarily~\cite{bob}
infer such a distant pion decay in terms of events which are {\underline{not coincident}}.

\subsection{From deformed momentum composition and relative spacetime locality to curved momentum space}
The two features I reviewed in the previous subsection, deformed laws of composition
of momentum and loss of absoluteness of distant coincidences of events,
must be clearly connected, but how? They must be connected because they are both
linked to translation transformations: a distant event for observer $O$ is
a nearby event for an observer $O'$ obtained from $O$ by a suitable
translation transformation; and momenta are charges associated with translational symmetries.
Essentially what was clearly missing was a satisfactory understanding
of some deformed notion of translation transformations
which would be compatible with the assumed deformations of Lorentz transformations.
A satisfactory way to address this challenge was offered in Refs.~\cite{principle,grf2nd},
within the novel ``relative-locality framework" centered on the geometry of momentum space.

For what concerns translations,
roughly speaking (detailed descriptions of these issues for translations
are in Refs.~\cite{principle,grf2nd,grbLeeLaurent,anatomy}),
the relative locality framework uses the modified conservation laws themselves
as generators of the translation transformations.
And it turns out~\cite{principle,grf2nd,grbLeeLaurent,anatomy},
as here illustrated in Fig.~2,
that these translations by themselves (even pure relative-locality-framework translations,
without any boosting) produce relative locality:
taken a pion-decay process described of course as a coincidence of events by the
nearby observer Alice this pion decay will be in general inferred in terms
of non-coincident events by a distant observer Bob (even if Bob is at rest with
 respect to Alice).

\begin{figure}[h!]
\begin{center}
\includegraphics[width=0.99 \textwidth]{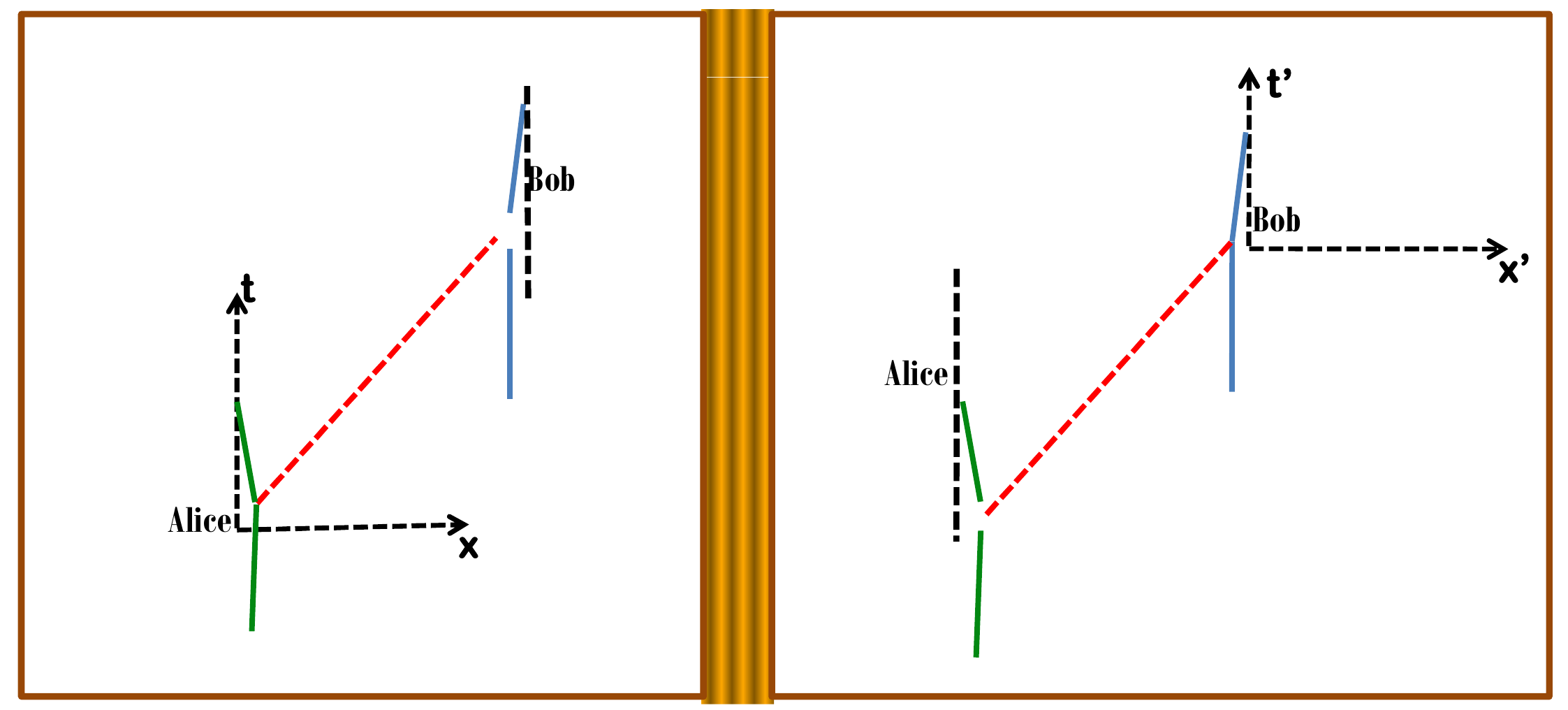}
\caption{I here illustrate a feature of relative spacetime locality that can arise
for suitable deformations of translation transformations, as shown in Refs.~[1,2].
I am imagining something like an atom being deexcited at Alice (near the origin
of  Alice's reference frame)
thereby exchanging a photon to a distant atom, which is then excited, at Bob
(near the origin
of  Bob's reference frame),
with Alice and Bob two distant observers in relative rest.
The left panel shows Alice's description of the coincidence of events she observes
for the first process and the lack of coincidence of events by which she would infer the
 second process.
The right panel shows Bob's description of the coincidence of events he observes
for the second process and the lack of coincidence of events by which he would infer the
first process.}
\end{center}
\end{figure}

In relation to Born's ``prophecy" it is noteworthy that all these properties
of the relative-locality framework can be described in terms of the geometry
of momentum space.
A  metric in momentum space $ds^{2}= g^{\mu\nu}(p) {\mathrm d}p_{\mu}{\mathrm d}p_{\nu}$
is used for writing the
energy-momentum on-shell relation,
$$m^{2} = D^{2}(p)~,$$
where $D(p)$ is the distance of the point $p_{\mu}$ from the origin $p_{\mu}=0$.
And a non-trivial affine connection  is needed for writing
 the law of composition of momenta,\footnote{The perspective on modifications of the on-shell
relation describing them in terms of an energy-dependent metric already surfaced
in the DSR literature before the advent of the relative-locality framework
(see, {\it e.g.}, Refs.~\cite{jurekDSmomentum,rainbowDSR}).
But only a metric (without a specification of affine connection)
does not make a geometry. The key conceptual ingredient of the relative-locality-framework
proposal is the description of deformed laws of composition of momenta
in terms of an affine connection on momentum space.}
$$\left(p \oplus_\ell q\right)_\mu \simeq p_\mu + q_\mu -
\ell \Gamma_\mu^{\alpha\beta}\, p_\alpha\,
q_\beta$$
where on the right-hand side the momenta are assumed to be small with respect
to $1/\ell$ and $\Gamma_{\mu}^{\alpha\beta}$
 are the connection coefficients evaluated at $p_\mu = 0$.

Another key achievement of the recent work on the relative-locality
curved-momentum-space framework is the ability to describe consistently
and systematically
interacting theories~\cite{principle,grf2nd,grbLeeLaurent,anatomy}
with deformed Lorentz symmetry, whereas earlier work with deformed Lorentz
symmetry was mainly confined to free theories. And this relative-locality
curved-momentum-space framework is not exclusively for theories
with deformed Lorentz symmetry: on the contrary it handles on the same footing
and with equal descriptive power both cases with geometry of momentum space
compatible with the introduction of DSR-deformations and
cases with geometry of momentum space that necessarily requires~\cite{goldenrule}
the introduction of a preferred class of observers.

\subsection{From curved momentum space to more boost-induced relative locality}
I have one more  step in this long logical chain from aspects of the quantum-gravity
problem through deformed boost transformations to curved momentum space.
The features of distant relative locality uncovered by DSR-deformed boosts
in Ref.~\cite{bob} were seen studying free particles.
Now that we have available the
description of interacting theories
given by the relative-locality curved-momentum-space framework
we can ask which additional features of relative locality are produced by deformed
boosts in an interacting theory. This is the issue
I studied in the recent Refs.~\cite{goldenrule,hopfhopf}.
An intriguing case is the one where the momentum-composition law
is
\begin{equation}
(p \oplus_\ell p^\prime)_1 = p_1 + e^{\ell p_0} p^\prime_1  ~,~~~
(p \oplus_\ell p^\prime)_0 = p_0 + p^\prime_0~.
\label{connectiontorsyALL}
\end{equation}
which within the relative-locality framework is viewed as a case with torsionful
momentum space.

A relativistic description of
interactions governed by the noncommutative composition law (\ref{connectiontorsyALL})
requires another conceptual leap in the logical chain I am describing in this section.
In our current theories (with linear momentum-composition law)
we can
simply impose that the boost of a two-particle\footnote{I focus here, to keep things short, on
two-particle events, which essentially describes the propagator and also describes
for example a transition from a $K_0$ to a $\bar{K}_0$.
The generalization to the more interesting case of events with more than two particles is in
Refs.~\cite{goldenrule,hopfhopf}.}
event $e_{k \oplus p}$  would be governed
by $N_{{k \oplus p}}= N_{[p]}+N_{[k]}$, {\it i.e.}
we assume implicitly that the boost of such a two-valent interaction can be
 decomposed
 into two pieces, each given in terms
of a boost acting exclusively on a certain momentum in the interaction.
This is our standard concept
 of ``total boost" generator obtained by combining trivially
the boost generators acting on each individual particle.
With noncommutative law of composition of momentum
this is in general no longer possible. In particular,
for the composition law (\ref{connectiontorsyALL})
one easily verifies that
$$[N_{[p]} + N_{[p^\prime]},
(p \oplus_\ell p^\prime)_\mu]\Big|_{(p \oplus_\ell p^\prime)_\mu=0} \neq 0~.$$
So if a relativistic description is at all available it will have to be
in terms of a deformed law of composition of boosts.

Looking for a suitable boost-composition law of this sort one quickly
realizes that, since the only deformation scale here available is an inverse-momentum scale,
the boost-composition law would have to involve momentum.
It is then not difficult\footnote{Guided by the line of reasoning I am here reporting
one can for example set up an {\it ansatz} for $N_{(p \oplus p^\prime)}$ given by a parametrized
formula involving as building blocks $N_{[p]}$, $N_{[p^\prime]}$ and the characteristic ``deformation
function" of (\ref{connectiontorsyALL}),
which is $e^{\ell  p_0}$. Testing the {\it ansatz} one then quickly finds
the option (\ref{boostgood}).} to establish what does work:
adopting the following boost-composition law
\begin{equation}
N_{(p \oplus p^\prime)} =N_{[p]}+ e^{\ell  p_0} N_{[p^\prime]}
\label{boostgood}
\end{equation}
one does find that
$[N_{(p \oplus p^\prime)},
(p \oplus_\ell p^\prime)_\mu]\Big|_{(p \oplus_\ell p^\prime)_\mu=0} = 0$.
This is easily verified:
\begin{eqnarray}
&& [N_{[p]}+ e^{\ell  p_0} N_{[p^\prime]},
p_0 + p^\prime_0 ]
= p_1 + e^{\ell p_0}p^\prime_1 = 0
\label{covathreezeroHHppkALL}
\end{eqnarray}
\begin{eqnarray}
&& [N_{[p]}+ e^{\ell  p_0} N_{[p^\prime]},
p_1 + e^{\ell p_0}p^\prime_1]
 = \frac{\ell }{2}
 \left(p_1 + e^{\ell p_0} p^\prime_1 \right)^2
  + \left(\frac{e^{2 \ell (p_0 + p^{\prime}_0) } -1}{2\ell}\right) = 0
\label{covathreeunoHHppkALL}
\end{eqnarray}
where for both these results I of course enforced on the right-hand side
the conservation law $p \oplus_\ell p^\prime = 0$ itself.

\section{A key example: $\kappa$-Poincar\'e-inspired momentum space}
In the previous section I discusses some new scenarios for relativistic kinematics.
We should expect that a full empowerment of research on these new (DSR-deformed)
scenarios for relativistic kinematics will also need the counterpart of
a corresponding description given in terms of symmetry algebras.
This balance is rather visible in special relativity, whose full understanding
requires combining the Poincar\'e symmetry algebra
and Einstein kinematics.
And we have one clear example (or at least a clear candidate example)
of notion of symmetry algebra which can accommodate all of the features
of relativistic kinematics highlighted in the previous section.
This is the $\kappa$-Poincar\'e Hopf-algebra framework
(see, {\it e.g.}, Refs.~\cite{lukie1,lukie2,majidruegg,kpoinap,majidREVIEW}).

\begin{figure}[h!]
\begin{center}
\includegraphics[width=0.96 \textwidth]{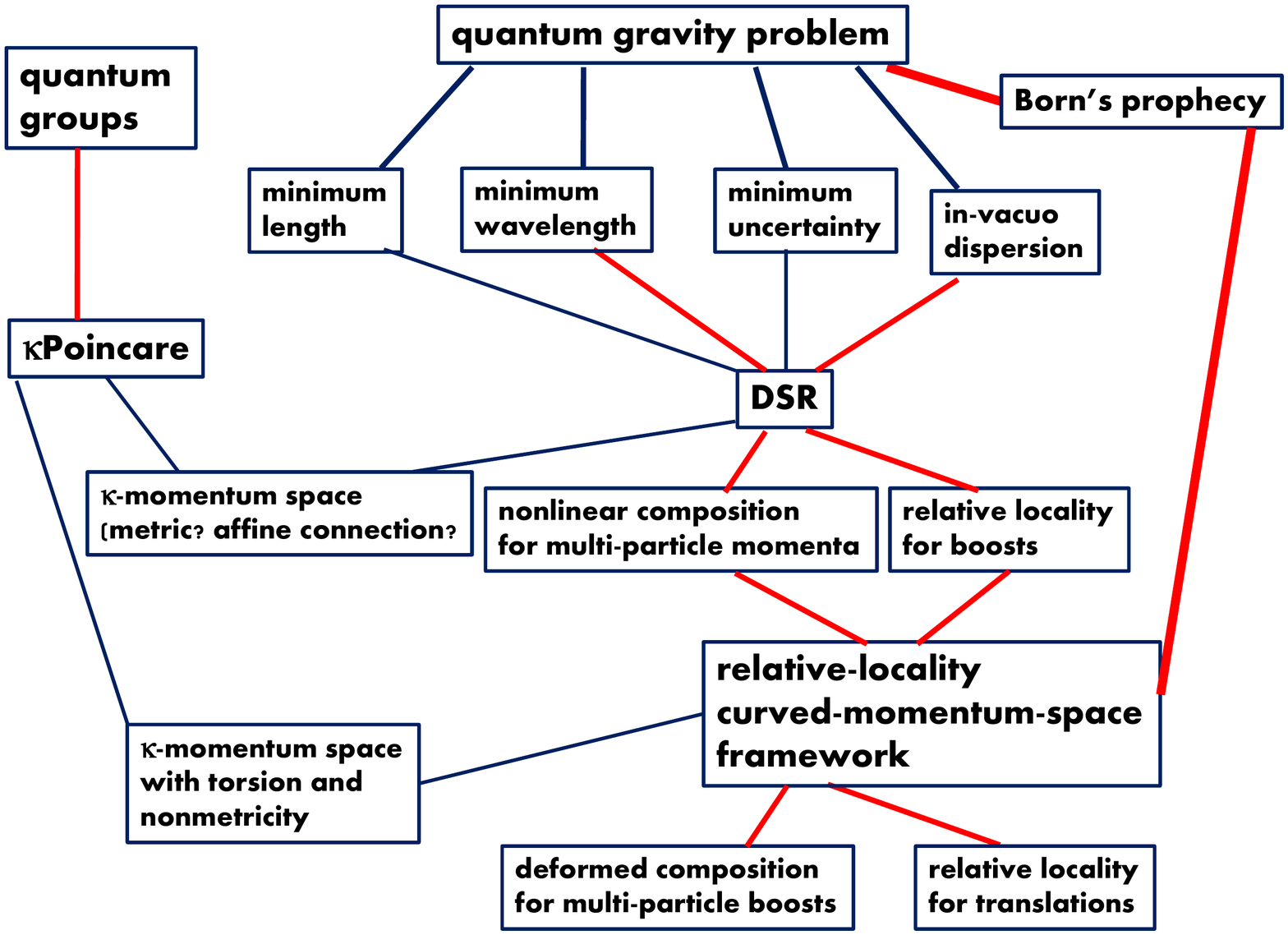}
\caption{I here show my preferred logical path from aspects of the quantum-gravity
problem to fulfilling Born's ``prophecy" of a curved momentum space for quantum gravity.
From notions such as a quantum-gravity-inspired ``minimum-wavelength principle"
one is guided toward DSR-deformations of Lorentz symmetry. From DSR deformations
one is led to nonlinearities in momentum space, including the idea
of nonlinear laws of composition of momenta. [But at this stage in the logical chain
these nonlinearties cannot yet be veiwed as ``geometric":
one could think of a momentum-dependent metric but
without a notion of parallel transport on the momentum space.]
And from DSR deformations one also manages to uncover a relativity of spacetime locality
produced by distant boosts.
Nonlinear laws of composition of momenta and relativity of spacetime locality
are then understood as two aspects of the same mechanism within the relative-locality
curved-momentum-space framework of Refs.~[1,2].
The idea of $\kappa$-Poincar\'e-inspired momentum spaces has proven to be a key theoretical
laboratory at each step in this logical chain.}
\end{center}
\end{figure}

In the previous section I have set the stage for making this point
by choosing accordingly my illustrative examples of the types of laws of relativistic
kinematics that could be adopted as a DSR scenario with a relative-locality
momentum-space description.
Essentially I now must only show the possible connection between relativistic
kinematics governed
by (\ref{disprelkappa}),(\ref{boostkappa}),(\ref{connectiontorsyALL}),(\ref{boostgood})
and the $\kappa$-Poincar\'e Hopf algebra.

This actually requires us to select a particular ``basis"~\cite{kpoinap}
for the $\kappa$-Poincar\'e Hopf algebra, which is the basis first introduced
in Ref.~\cite{majidruegg}. A relation to  (\ref{disprelkappa})
and (\ref{boostkappa})
is evident in the commutators between $\kappa$-Poincar\'e generators
of boost and momenta in that basis (again specializing to the 1+1-D case):
\begin{equation}
[{\cal N}, {\cal P}_0] =  {\cal P}_1 ~,~~~
[{\cal N}, {\cal P}_1] = \frac{e^{2 \ell {\cal P}_0} - 1}{2\ell}  + \frac{\ell}{2} {\cal P}_1^2~.
\label{boostkapparight}
\end{equation}
from which one finds the following deformed mass Casimir:
\begin{equation}
\cosh (\ell {\cal P}_0) - \frac{\ell^2}{2} e^{-\ell {\cal P}_0} {\cal P}_1^2
\label{casimirkappa}
\end{equation}

In relation to (\ref{connectiontorsyALL}),(\ref{boostgood}) a connection
is visible with the co-algebra (co-products) structure of
the $\kappa$-Poincar\'e Hopf algebra in the basis of Ref.~\cite{majidruegg},
which read
\begin{equation}
\Delta({\cal P}_0) = {\cal P}_0 \otimes 1 + 1 \otimes {\cal P}_0~,~~~
\Delta({\cal P}_1) = {\cal P}_1 \otimes 1 + e^{\ell {\cal P}_0} \otimes {\cal P}_1~,
\label{coproP}
\end{equation}
\begin{equation}
\Delta({\cal N}) = {\cal N} \otimes 1 + e^{\ell {\cal P}_0} \otimes {\cal N}~.
\label{coproN}
\end{equation}

Starting from these connections here highlighted I think it is legitimate to
expect that the $\kappa$-Poincar\'e Hopf algebra will continue to be
a precious ``theoretical laboratory" for the future development
of the quantum-gravity research line that I described schematically in the preceding section.
It has been so far the most valuable ``theoretical laboratory" of this sort
(see Fig.~3).
We now are in better position for seeing other similarly useful examples of
novel symmetry-algebra structures potentially relevant for
the development of Planck-scale-deformed kinematics. I conjecture for example
that something similar to (but ultimately different from)
the  $\kappa$-Poincar\'e Hopf algebra will be needed in order to
fully investigate some of most recent ideas that emerged on the deformed-kinematics
side, such as the notion of ``dual-gravity lensing"~\cite{grbLeeLaurent,transverse}

\section{Implications for quantum gravity}
Relative locality is a spacetime feature which we can (and must) consider
once we appreciate that spacetime is redundant. It is a way by which observers can
still abstract a rather conventional spacetime picture, but since those pictures
are only collections of inferences about a redundant structure, the spacetime abstraction
of one observer does not need to match the spacetime abstraction of another observer.
Spacetime points (coincident events) for one observer do not need to be spacetime points
for another observer.

I have discussed a path toward the adoption of relative locality which is based on
a perspective on the quantum-gravity problem. This is a path going through deformed Lorentz
symmetry and deformed relativistic kinematics. Evidently if there is a deformation
of relativistic kinematics it will have very significant implications on the nature
of the quantum-gravity realm.

In this closing section I want to contemplate other sorts of implications
for quantum-gravity research that could come from the realization of the
redundancy of spacetime. I stress again that this is the main objective of
these notes: the question of whether spacetime ``exists" or ``does not exist",
and even to a large extent the issues of the redundancy of spacetime, falls well
beyond the boundary of science, so it is of no interest to me, but an appreciation
of the redundancy of spacetime in physics (in the sense discussed in the previous sections)
may be relevant for building intuition for how future theories may be structured.
Relativity of spacetime locality is an example of this, but other even wilder
pictures should in principle be considered once we appreciate the redundancy of spacetime.

\subsection{Implications for early-Universe cosmology}
One of the limitations of the way humans learn physics is that our condition
usually puts us in the position of going ``upstream". A good example
is the program of ``quantization of theories": from Nature's perspective
theories start off being quantum theories and happen to be amenable
to description in terms of classical mechanics only in some peculiar limiting cases,
but our condition is such that we experience more easily those limiting
cases rather than the full quantum manifestation of the laws. So we started
with, {\it e.g.}, Maxwell's theory and then we learned about QED, a ``quantization"
of Maxwell's theory. The logical path would have been much clearer (or at least clear
much sooner) if we had the luxury of going downstream, first experiencing QED and
then discovering Maxwell's theory as a special limit of QED.

In light of the observations here reported we must infer that our ``scientific
relationship" with spacetime is another example of going upstream.
Spacetime is a feature (an abstraction characterizing) our primitive measurements
which are timed particle detections.  I do not see any way to introduce
spacetime operatively without clocks and detectors. And yet it is standard to
develop theories by introducing the spacetime picture as first ingredient,
than introducing a long list of formal properties of fields (or particles)
in that spacetime, and only in the end (for those brave enough to get that far)
we worry about actually having detectors and clocks in our theory.\\
This works. It worked so far. It works in the same sense (and for the same reasons)
of the success of the program of ``quantization of theories".
But this luxury of going upstream might be lost at some point.

An example is early-Universe cosmology: our current models, even for sub-Planckian times,
assume a spacetime picture. In light of the line of reasoning I have here advocated
I would be very curious to see proposals of a description of the early Universe
which initially (for a finite ``duration") has no spacetime.
At Planckian energies, and without any macroscopic
measuring device (detectors, clocks) to speak of, we might have already processes
and yet not being provided the luxury of abstracting a spacetime, not even a spacetime
affected by relative locality.

\subsection{Implications for ``emergent gravity"/``emergent spacetime"}
For reasons similar to the ones I just stressed in relation to the early Universe,
awareness of the redundancy of spacetime may be valuable for the fashionable
idea of ``emergent spacetime" and ``emergent gravity".
The point here is in a subtle change of intuition: with ``emergent spacetime"
one is prepared to describe spacetime as not being ``fundamental", typically
being a structure that takes the shape we are familiar with only for low-energy probes
and instead has a different structure ``fundamentally". The line of reasoning I
hare advocated is such that spacetime might ``emerge" in a rather different sense.
It may well be that the abstraction of a spacetime admits
a deeper level of abstraction, but this is not the key ``emergence issue"
we would worry about from a spacetime-redundancy perspective:
we would want to know how and when (under which conditions) we are afforded
the abstraction of a spacetime, and what replaces reliance on that abstraction
in regimes where it does not apply. And is it not the case that we should only
expect a spacetime picture to arise/emerge
when actual physical/material clocks and detectors
can be properly contemplated?

\subsection{Triviality of a holographic description of black holes}
The indirect evidence we have concerning the possibility to describe holographically black holes
appears to be puzzling if one conceptualizes the physics of black holes to be ``contained"
in the 4D spacetime region with boundary the black-hole horizon.
According to the perspective I here advocated one should conclude that describing the
physics of black holes as contained
in the 4D spacetime region with boundary the black-hole horizon
is very naive. In the situation we colloquially describe as being in presence of
a black hole we are ultimately describing emission/detection correlations among
a collection of emitter/detectors at (or all around) the horizon of the black hole.
The idea of ``Hawking radiation" for example pertains to the cause/effect correlations
that could be found when one of these emitter/detectors emits a particle ({\it e.g.} ``toward
the black hole") and when one of these emitter/detectors detects a particle (a particle of
Hawking radiation in the most interesting cases).
The presence of an horizon should be more carefully described not in terms of the
structure of the abstracted spacetime but rather in terms of the network
of emitter/detectors that can be setup.
Ultimately horizons are not to be described in terms of properties of the (abstracted)
spacetime, but rather as limitations to signal exchange among (real, physical) emitter/detectors.
There is no use for emitter/detectors
to be contemplated inside the black hole: there is an (at least one-way) horizon between
emitter/detectors inside the black hole and emitter/detectors outside the black hole.

From all this we deduce that the description of black holes is necessarily holographic,
since it pertains to a distribution of emitter/detectors which is 2+1-dimensional.
As a matter of fact even for classical black holes, where formally it seems we can
introduce a non-holographic description of black holes, by assigning fields at each point
of spacetime, including the interior of the black hole, the description of black holes is ultimately
holographic: the values of the fields inside the black hole are not observable.
Any theory which describes classical gravity exactly like Einstein gravity for the outside of
black holes but gives a different description of the inside of black holes
is to be described as exactly equivalent scientifically to Einstein gravity, since
no measurement result could describe the difference between these two theories.
It is rather intriguing that we
have some evidence of the fact that the quantum description of black holes
may be explicitly holographic (rather than being implicitly holographic like its classical-limit
counterpart).
But it should be acknowledged that the physics of black holes is inevitably
holographic: the formalization may or may not be explicitly holographic but the
physics is indeed inevitably holographic.

\subsection{New paths to quantum gravity}
An example of new path toward a solution of the quantum-gravity problem
that can be inspired by awareness of the redundancy of spacetime can take as
starting point the novel relative-locality curved-momentum-space framework.

In proposing this picture, Freidel, Kowalski-Glikman, Smolin and I
adopted the standpoint~\cite{principle,grf2nd} that theories are formulated
on momentum space, which works well with the redundancy of spacetime.
I here just want to stress that in the relative-locality curved-momentum-space
framework Newton constant, $G_N$, may be redundant. This is simply because
one needs~\cite{principle,grf2nd} the Planck scale as scale of geometry
of momentum space, and geometry of momentum space is primitive in the relative-locality
framework. So rather than viewing the Planck scale as derived from Newton constant
(plus Planck constant ``$\hbar$" and speed-of-light scale ``$c$"), within the relative-locality
framework it is more natural to view Newton constant as derived from the Planck scale:
$$G_N = \frac{\hbar^2 c^5}{E_P^2}$$

Combining awareness of the redundancy of spacetime with the
perspective of the relative-locality curved-momentum-space framework, one can
envisage profoundly new ways for describing ``spacetime physics".
As an example of what this might lead to,
I observe in Fig.~4 that one could contemplate 
 a new version of the Bronstein cube.

\begin{figure}[h!]
\begin{center}
\includegraphics[width=0.62 \textwidth]{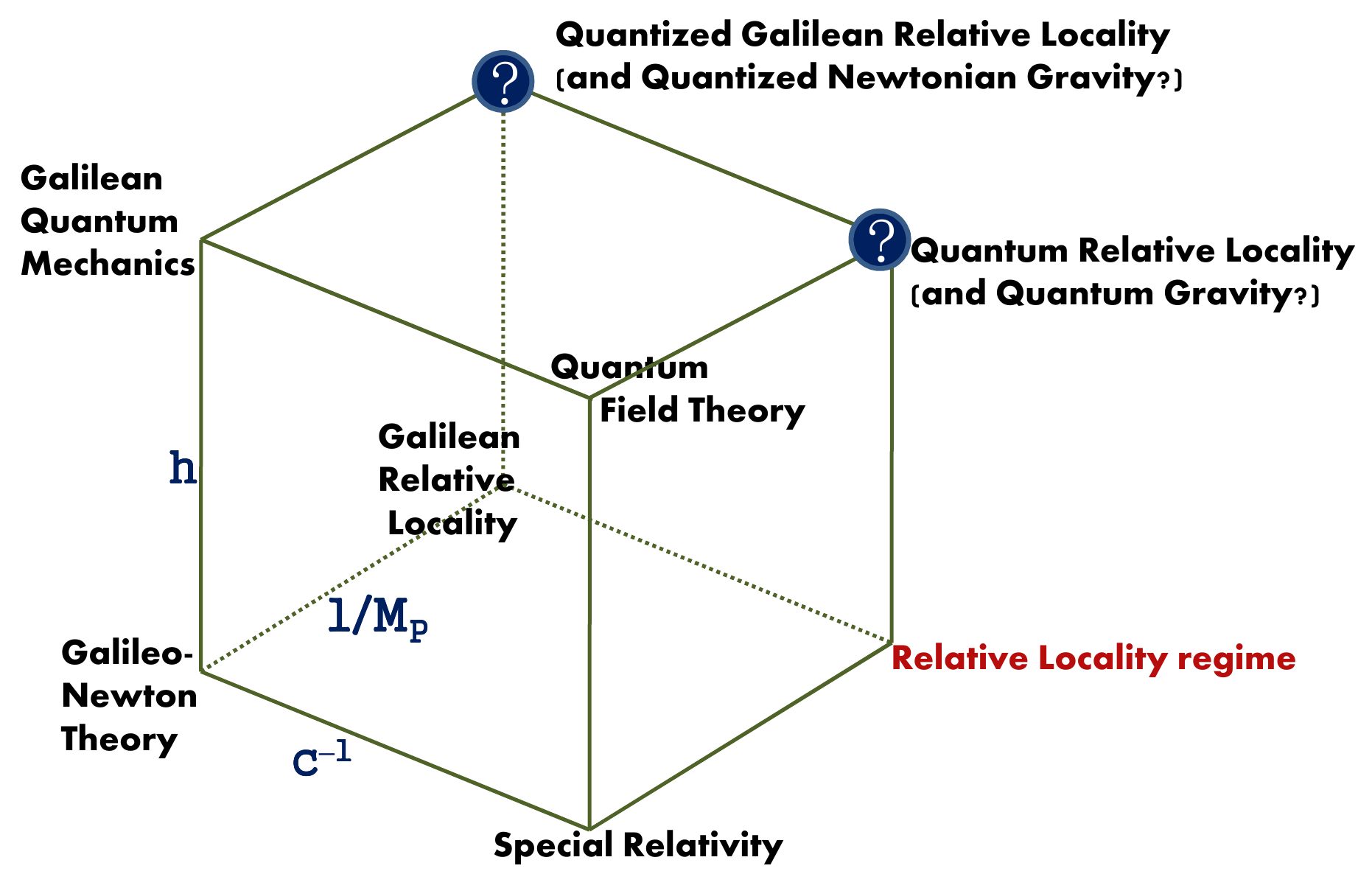}
\caption{If the Planck scale is primitive (together with $\hbar$ and $c$)
while Newton constant is derived, the Bronstein cube would have to be redrawn as here shown.
I am assuming that the relative-locality regime is described by the relative-locality
framework
or something similar to it. This may also
suggest that
that the theory encompassing all these regimes (``quantum gravity") could be obtained
as a quantum theory on the relative-locality momentum space.}
\end{center}
\end{figure}

\newpage
From this perspective
I am particularly intrigued by the fact that we are used to thinking
that spacetime curvature produces noncommutativity of the momenta/translation generators
(readers unfamiliar with this may think in particular of de Sitter spacetime and its symmetry
algebra).
So one could perhaps attempt to codify the geometry of the (inferred/redundant) spacetime
within rules of noncommutativity of momenta at a given observer/detector. And something like
Einstein's equations for the dynamics of geometry of spacetime could perhaps also
be written in terms of laws for the noncommutativity of momenta for different observers/detectors.

And it could be something even wilder than this. Gravity could be a purely quantum-mechanical effect:
 on the basis of $G_N = \hbar^2 c^5/E_P^2$
one could be tempted of viewing Newton constant as a feature of the structure at
order $\hbar^2$ of quantum theories on a curved momentum space.

\section*{Acknowledgements}
I am very grateful to the organizers for the privilege
of speaking on these subjects at the XXIX Max Born Symposium,
held at Wroclaw's Institute of Theoretical Physics (located on plac Maxa Borna)
where many of the results on $\kappa$-Poincar\'e were obtained.
And the fact that we were also celebrating Jerzy Lukierski's birthday
rendered the meeting an even more special event.
It was fun to collaborate and exchange ideas with Jerzy
over the years~\cite{gacLUKIE1997,gacLUKIE1999,gacLUKIE2001},
and I want to take this opportunity
also to express gratitude to Jerzy for his kindness during our first chance
encounter at a meeting in Alushta (Ukraine) in 1996. I was then rather young and
completely unknown to the quantum-spacetime/quantum-gravity community,
but I was in the process of  transitioning  from a mostly
field-theory/particle-physics focus (with quantum gravity as a
hobby~\cite{gacmpla1994,gacmpla1996})
to having quantum gravity as my main research interest.
After his talk at Alushta on $\kappa$-Poincar\'e/$\kappa$-Minkowski,
Jerzy was extremely generous
in attending to my curiosity for this formalism which has so much fitting with
my intuition~\cite{gacmpla1994,gacmpla1996} for quantum spacetime.
When these days it happens that a young physicist asks me some quantum-gravity questions
I often remember Jerzy's example, and try to follow it.

\end{document}